\documentclass[aps,prl,twocolumn,showpacs,amsmath,amssymb]{revtex4}

\usepackage{graphicx}
\usepackage{dcolumn}
\usepackage{bm}
\usepackage{textcomp,mathcomp}

\begin{document}
\title{Condensate splitting in an asymmetric double well for atom chip based sensors}

\author{B. V. Hall}
\email{brhall@swin.edu.au}
\author{S. Whitlock}
\author{R. Anderson}
\author{P. Hannaford}
\author{A. I. Sidorov}

\affiliation{ARC Centre of Excellence for Quantum-Atom Optics and
\\Centre for Atom Optics and Ultrafast Spectroscopy, Swinburne
University of Technology, Hawthorn, Victoria 3122, Australia }

\date{\today}

\begin{abstract}

We report on the adiabatic splitting of a BEC of $^{87}$Rb atoms by
an asymmetric double-well potential located above the edge of a
perpendicularly magnetized TbGdFeCo film atom chip. By controlling
the barrier height and double-well asymmetry the sensitivity of the
axial splitting process is investigated through observation of the
fractional atom distribution between the left and right wells.  This
process constitutes a novel sensor for which we infer a single shot
sensitivity to gravity fields of $\delta g/g\approx2\times10^{-4}$.
From a simple analytic model we propose improvements to chip-based
gravity detectors using this demonstrated methodology.
\end{abstract}


\maketitle

Atom interferometry techniques have been successfully applied to the
precision sensing of inertial forces, with demonstrated measurement
of the Earth's rotation rate and the acceleration due to the Earth's
gravity \cite{Pet99a,Gus97}.  More recently with the advent of atom
chips physicists are seeking to exploit atom interferometric methods
using trapped samples of ultra-cold atoms and Bose-Einstein
condensates (BEC) \cite{Wan05,Sch05}. While these trapped atom
interferometers are in their infancy, a number of pioneering
experiments have already utilized the inherent low energy scale of
ultra-cold atoms and BECs as probes of external phenomena. Such
experiments include using BECs/ultra-cold atoms to probe the
magnetic potential landscape above conducting micro-wires
\cite{Wil05a} and magnetic films \cite{Whi06}; measuring the
electric field distribution on charged micro-wires \cite{Wil05b};
quantifying the amount of near-field radiation due to Johnson noise
in conductive and dielectric surfaces \cite{Jon03,Sin05b}; and using
the dynamical behavior of a condensate to precisely measure the
Casimir-Polder force as a function of distance \cite{Har05}. In
addition trapped bosonic $^{88}$Sr atoms at nano-Kelvin temperatures
have recently been used to measure gravity with a estimated
sensitivity of $\delta g/g\approx4\times10^{-5}$ \cite{Fer06}.

\begin{figure}
\includegraphics{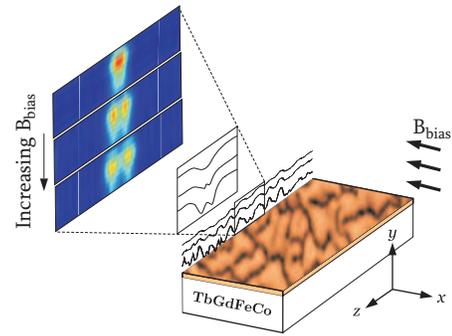}
\caption{\label{fig:chip}(Color online) Schematic showing the
corrugated potential above the edge of the TbGdFeCo film. A double
well can be used to axially split a BEC by increasing B$_{bias}$.
The corrugated potential is due to the presence of inhomogeneity in
the magnetic film.}
\end{figure}

In this Letter we report on a novel sensor which utilizes adiabatic
axial splitting of a BEC in an asymmetric double well. This
potential is created above a perpendicularly magnetized TbGdFeCo
film \cite{Hal06} which has previously been used to study the
influence of magnetic film inhomogeneity on ultra-cold atoms
\cite{Whi06}.  We show that while cloud fragmentation was initially
reported as a deleterious effect for atom chips \cite{For02,Est04},
our unique potential landscape can be utilized to realize a
precision sensor which measures potential gradients. Time dependent
magnetic fields in combination with pulsed radio frequency (rf)
spectroscopy and absorption imagining are used to characterize the
double-well separation $\lambda$, the barrier height $\beta$ and the
asymmetry $\Delta$ as a function of trap-surface distance. Results
are then presented for adiabatic BEC splitting experiments where the
fractional atom number distribution is used as a measure of the
double-well asymmetry to much higher precision than that obtained
with rf spectroscopy.  A simple analytic model is presented to
determine the sensitivity of this technique and its dependence on
atom number, trap parameters and atomic properties. Finally we use
this model to propose improvements to atom chip sensors based on
this method.

A detailed description of the apparatus, including experiments
performed with a BEC on a perpendicularly magnetized film atom chip,
have been described elsewhere \cite{Hal06,Whi06}.  In brief, our
atom chip consists of a uniformly magnetized TbGdFeCo film on a
0.3~mm thick glass slide which is epoxied to a machined Ag foil
structure. The Ag structure allows currents to be passed in U- and
Z-shape paths for surface magneto-optical trap (MOT) and
Ioffe-Pritchard trap geometries \cite{Fol00}.  A uniform magnetic
field $B_{bias}$ combines with the magnetic field above the film
edge to provide radial confinement while a pair of end wires provide
axial confinement. A small amount of inhomogeneity in the TbGdFeCo
film results in a corrugation of the axial potential with a period
of about 390~\textmu m. This corrugating potential introduces
fragmentation to the ultra-cold atom cloud which becomes more
pronounced as the cloud approaches the surface. The measured
potential is depicted in Fig.~\ref{fig:chip} along with a schematic
of the region of interest, where the double well is located with its
center at z=0. This double well arises from higher frequency
components of the magnetization inhomogeneity exerting a larger
influence as the film trap is brought closer to the surface with
increasing $B_{bias}$ in the x-direction.

\begin{figure}
\includegraphics{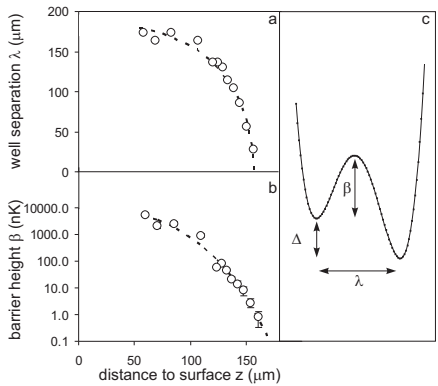}
\caption{\label{fig:char}Characterization of the double well as a
function of trap-surface separation is performed using two component
clouds. The dashed line in (a) and (b) are to guide the eye. The
well separation $\lambda$, barrier height $\beta$ and trap asymmetry
$\Delta$ are shown schematically in (c).}
\end{figure}

Since the magnetic topography is created by random magnetic
inhomogeneities present in the film, a systematic characterization
of the double-well system is required. To achieve this a partially
condensed cloud of 10$^5$ $^{87}$Rb atoms in the $F=2, m_{F}=2$
state is first prepared in a single-well potential located
170~\textmu m from the surface.  Over 500~ms $B_{bias}$ is ramped to
the desired value at which point the trap is rapidly switched off.
The spatial distribution is determined via absorption imaging after
a 2~ms expansion time. The potential is then mapped by fitting
standard Thomas-Fermi/Gaussian distribution functions which are
scaled by the cloud temperature and chemical potential $\mu$
\cite{Wil05a}. From these measurements the separation $\lambda$ and
the barrier height $\beta$ of the double-well system are determined
for a range of trap heights, (Fig.~\ref{fig:char}).  Faster
splitting times are also used to induce dipole mode oscillations in
the condensate to measure both the axial and radial trap frequencies
\cite{Hal05}. Remarkably, the left and right wells have trapping
frequencies which are identical to within a few percent.

Another parameter of interest depicted in Fig.~\ref{fig:char}c is
the asymmetry $\Delta$ between the left and right wells. A non-zero
$\Delta$ results in a marked lop-sidedness in condensate
distribution for splitting experiments when the barrier height is
smaller than the chemical potential ($\beta<\mu$). Similar
observations involving the radial splitting of a thermal cloud in a
double-well potential have been reported previously \cite{Est05}.
Prior to this investigation the apparatus was mounted on a
$\it{floating}$ vibration-isolating optical table. Systematic drifts
in the tilt of the chip were observed throughout the day due to
changes in line pressure feeding the isolating legs. Operating the
table as a rigid structure eliminated this effect; however a slight
tilt of the atom chip with respect to gravity still remains for
identical end wire currents. This gravitational gradient is canceled
in the above experiments by applying a small magnetic field gradient
$\frac{\partial B_{z}}{\partial z}$ which is controlled by an end
wire current imbalance $\pm\delta I$. For two infinitely thin
parallel wires propagating current in the $\widehat{x}$ direction,
separated by a distance $\it{d}$, the gradient along $\it{z}$ at
height $\it{y}$ above the midpoint between these wires is given by

\begin{figure}
\includegraphics{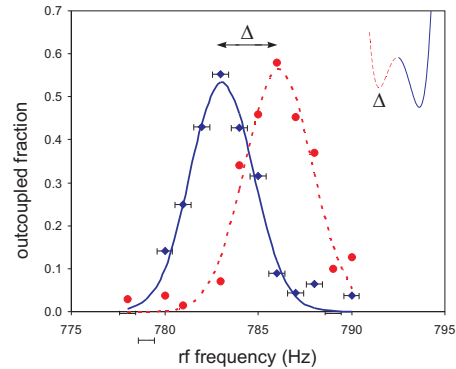}
\caption{\label{fig:calibrate}(Color online) Results of pulsed radio
frequency spectroscopy, performed on a condensate which has been
split symmetrically and then exposed to a large asymmetry.  The
shift in the output coupled spectra yields a measure of $\Delta$.}
\end{figure}
\begin{equation}
\label{eq:dBdy} \frac{\partial B_{z}}{\partial z}~=~\frac{8~\mu_{0}~
\delta I}{\pi}~\frac{d~y}{(d^2+4~y^2)^2}~~~~~~~~~~Tm^{-1}
\end{equation}

Equation \ref{eq:dBdy} illustrates that for a fixed height and
separation the magnetic field gradient is proportional to $\delta
I$. To circumvent finite size effects and imbalances in the machined
end wire uniformity a method of calibrating $\Delta$ $\it{in~situ}$
was developed. This entails adiabatically splitting a condensate in
a $\it{symmetric}$ double well with barrier height $\beta>\mu$ such
that no spillage can occur between the left and right wells. A large
asymmetry is then applied over 500~ms by tuning the end wire current
imbalance $\delta I$. Next a low intensity 40~\textmu s pulse of
fixed frequency rf radiation (Rabi frequency
$\Omega/2\pi$=~0.43~kHz) is introduced. This outcouples atoms from
either well in a magnetic energy selective manner
(h$\nu$=$g_Fm_F\mu_B|B|$) and this outcoupled fraction is measured
by absorption imaging. Figure~\ref{fig:calibrate} shows the results
of this spectroscopic technique where the separation of the
left/right well distributions is a measure of the asymmetry
$\Delta$.  This calibration process was repeated for various $\delta
I$ values yielding $\Delta$= 2.0(1)~Hz/mA.

We now focus on the dynamical splitting of a condensate in a double
well with asymmetry.  To remain in the ground state of the potential
the splitting time $t_{s}$ must be greater than $\Delta^{-1}$ thus
ensuring adiabaticity of the splitting process. This condition
establishes a lower bound for the resolvable asymmetry of this
method which can be related to the following experimental
limitations on $t_{s}$.  The longest splitting time is determined by
background gas collisions ($\tau_{Rb}\approx$~100 s) which decrease
the sensitivity of the measurement by reducing the signal to noise
ratio by $e^{\frac{-t_s}{2\tau_{Rb}}}$.  A more relevant limitation
is the condensate lifetime $\tau_{C}$ which can range from 20~ms to
several seconds.  Processes which shorten $\tau_{C}$ above atom
chips include spin flip loss near surfaces \cite{Hen99}, parametric
heating due to technical noise \cite{Hen03}, density dependent
collisional loss and heating mechanisms \cite{Str06}.

The magnetic trapping potential generated by the TbGdFeCo film is
well suited for minimizing these limitations. The double well exists
far from the chip surface (170~\textmu m) which, in conjunction with
the reduced thickness of the TbGdFeCo-Cr-Au material (1~\textmu m),
significantly lowers the spin-flip loss rate. In addition, due to
the nature of the perpendicularly magnetized film, technical noise
due to current fluctuations are not present on the structure that
generates the corrugated potential. Moreover the radial and axial
trapping potentials are relatively weak and the number of condensed
atoms is kept low to avoid high density loss and heating processes.
With these conditions we have measured heating rates for small
thermal clouds in single-well magnetic film traps as low as 3~nK/s
and condensate lifetimes of 5~s have been observed without the
application of a radio-frequency shield. In future experiments this
lifetime may be further improved by preparing the condensate in the
($F=1, m_{F}=-1$) ground state.

\begin{figure}
\includegraphics{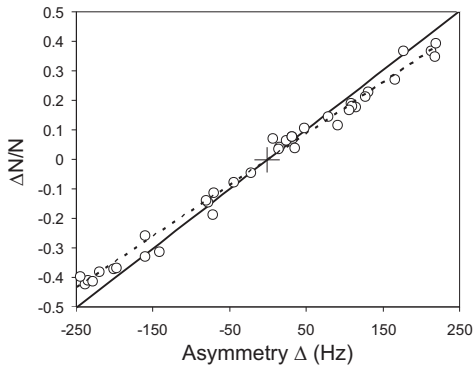}
\caption{\label{fig:frac}Fractional number difference versus the
double-well potential asymmetry. The asymmetry is varied by randomly
changing the end wire current imbalance $\delta I$.  The
experimental data points (open circles) and the line of best fit
(dashed line) compare well with the simple analytic result (solid
line).}
\end{figure}

With $\Delta$ calibrated the sensitivity of the splitting process
for small values of $\Delta$ and a large splitting time $t_{s}$ is
examined. A small condensate of $4.6\times10^{4}$ atoms is prepared
in a single well 170~\textmu m from the film surface. $B_{bias}$ is
ramped between 0.130 and 0.135~\textmu T over 2~s, producing a
tailored double well 155~\textmu m from the surface with a well
separation $\lambda\approx$~70~\textmu m. At this position
$\beta\approx\frac{1}{4}\mu$ and the chemical potential of the right
($\mu$) and left ($\mu'$) wells is given by $\mu=\mu'+\Delta$. The
asymmetry is set by the end wire current imbalance which ranges
between $\pm$~117~mA over the course of these experiments. At the
end of the adiabatic splitting process $B_{bias}$ is rapidly
increased over 10~ms to further separate each well thus enhancing
the spatial discrimination of the imaging process. All trapping
fields are then turned off and each cloud expanded for 2~ms prior to
taking an absorption image. A CCD camera captures the image with a
resolution of 3.5~\textmu m per pixel. This experimental cycle takes
30~s and is repeated varying only the current imbalance $\delta I$
to collect a complete data set. The atom number difference between
the left and right wells is then normalized by the sum $N_L+N_R$
yielding $\Delta N/N$ for each $\delta I$ measurement. Figure
\ref{fig:frac} shows the results from 35 individual $\delta I$
measurements collected randomly in time.  From the distribution of
the data we estimate that a measurement of $\Delta$ can be performed
with a single shot sensitivity of 16~Hz where the dominant source of
noise has been attributed to shot-to-shot condensate number
fluctuation at the 3 percent level.  This resolution is
significantly better than that obtained with rf-spectroscopy which
is typically limited by fluctuating background magnetic field noise
at the $0.1$~\textmu T level. Moreover, since the total potential
asymmetry is measured, $\Delta$ is also sensitive to non-magnetic
field sources such as gravity fields or light shifts. Although a
magnetic field gradient is used above it is conceivable that the
same measurements could be performed by tilting the chip or by
introducing a gravitational field from a test mass. Using the
70~\textmu m separation we infer a single shot sensitivity to a
gravity field of $\frac{\delta g}{g}=2\times10^{-4}$. At present
this accuracy surpasses interferometric experiments with trapped
atoms which appear to be limited due to phase decoherence for
splitting times greater than a few milliseconds \cite{Sch05}.

\begin{figure}
\includegraphics{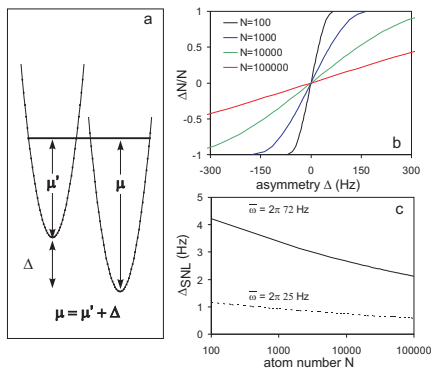}
\caption{\label{fig:theory}.  The asymmetric double well is
represented by two identical harmonic traps offset by $\Delta$ with
a ground state which obeys $\mu=\mu'+\Delta$ (a). From this model we
can extract the dependence of the fractional number difference on
$\Delta$ for increasing atom number (b).  In the standard quantum
limit the sensitivity improves as $N^{-0.1}$ as shown for two trap
conditions (c).}
\end{figure}

To provide a further understanding of this experiment a simple model
is proposed whereby the double-well system is represented by two
uncoupled 3-dimensional harmonic wells with ground states shown in
Fig.~\ref{fig:theory}a.  Using the Thomas-Fermi approximation for a
harmonic trap the chemical potential $\mu$ is given by,

\begin{equation}
\label{eq:chempot} \mu(Hz)=\frac{\bar{\omega}}{4\pi}\left(
\frac{15N_{R}a}{a_0}\right)^{\frac{2}{5}}=\frac{\bar{\omega}}{4\pi}\left(\frac{15N_{L}a}{a_0}\right)^{\frac{2}{5}}+\Delta
\end{equation}

where $N_{L}$ and $N_{R}$ are the number of atoms in the left and
right wells respectively, $\bar\omega$ is the geometric mean of the
trap frequencies, $a$ is the atom scattering length and $a_{0}$ is
the ground state harmonic oscillator length. Algebraic manipulation
and a first order expansion about $\Delta$~=~0 yields the following
relation for the fractional number difference between the left and
right wells

\begin{equation}
\label{eq:DNoverN} \frac{\Delta N}{N}\approx~1.65~\frac{\Delta}{c
N^{\frac{2}{5}}}~~~~~~~~c=\frac{\bar{\omega}}{4\pi}\left(\frac{15a}{a_0}\right)^{\frac{2}{5}}
\end{equation}

Inserting experimental parameters into Eq.~\ref{eq:DNoverN} yields
the solid line in Fig.~\ref{fig:frac}. This shows good agreement
with the data considering the simplicity of the model. To
investigate the limits of this technique $\frac{\Delta N}{N}$ was
evaluated using Eq.~\ref{eq:chempot} for several N values
(Fig.~\ref{fig:theory}b). Decreasing atom number reduces $\mu$ such
that smaller asymmetries yield larger $\frac{\Delta N}{N}$. However
in the shot-noise limit the fractional number difference can be
measured more accurately with increasing atom number as
$\sigma_{\Delta N/N} \propto N^{-1/2}$. These two opposing behaviors
are evident in Fig.~\ref{fig:theory}c which shows that the
shot-noise limit sensitivity depends only weakly on atom number
($N^{-0.1}$). A more appropriate strategy to improve the measurement
sensitivity is to decrease $\bar\omega$ thus relaxing the potential
and lowering $\mu$ accordingly.  Another approach which lends itself
to the atom chip concept is to micro-fabricate multiple ($\times n$)
double-well potentials on a single chip thus gaining $\sqrt{n}$
enhancement per measurement cycle.  Reasonable parameters for such a
chip (n=100, N$_{L,R}$~=~2000 atoms/well and
$\bar{\omega}=2\pi\times$12~Hz) yield a possible single shot
sensitivity of $\Delta_{SNL}\approx$~0.04~Hz or $\frac{\delta
g}{g}=5\times10^{-7}$ for a well separation $\lambda$=~70~\textmu m.
While this sensitivity appears low compared to state-of-the-art
gravimeters, it already falls into the range of interest for mineral
prospecting while offering an atom chip sized package which is
ultimately scalable \cite{Pet99b}.

In conclusion, a corrugated potential resulting from random
magnetization inhomogeneities has been used to investigate the
dynamical transfer of a condensate from a single to a double-well
potential.  Using radio frequency spectroscopy and absorption
imaging techniques this potential was characterized.  The
sensitivity of the fractional number difference to a magnetic field
gradient induced asymmetry was then investigated. A simple model
predicting the sensitivity of this technique agrees well with
observations and proves useful in motivating additional improvements
to the gravity field sensitivity of future experiments.

\begin{acknowledgments}
This project is supported by the ARC Centre of Excellence for
Quantum-Atom Optics and a Swinburne University Strategic Initiative
grant.
\end{acknowledgments}

\bibliography{hall_doublewell}

\end{document}